\documentclass[showpacs,twocolumn,pre,floatfix]{revtex4}
\usepackage{psfrag,epsfig,amsfonts,amssymb,amsmath}
\usepackage{dcolumn}

\newcommand{\bx}{{\bf x}}
\newcommand{\bu}{{\bf u}}
\newcommand{\bJ}{{\bf J}}
\newcommand{\bE}{{\bf E}}
\newcommand{\NA}{N_{\mathrm{A}}}
\newcommand{\kB}{k_{\mathrm{B}}}
\newcommand{\amin}{a_{\mathrm{min}}}
\newcommand{\dr}{\frac{\mathrm d}{\mathrm dr}}
\newcommand{\sign}{{\mathrm{sign}}}
\begin{document}

\title{On the reluctance of a neutral nanoparticle to enter a charged pore}

\author{Sebastian Getfert}
\author{Thomas T\"ows}
\author{Peter Reimann}
\affiliation{Fakult\"at f\"ur Physik, 
Universit\"at Bielefeld,  33615 Bielefeld, Germany}

\begin{abstract}
We consider the translocation of a neutral 
(uncharged) nanoparticle through a pore
in a thin membrane with constant surface 
charge density.
If the concomitant Debye screening layer is
sufficiently thin, the resulting forces experienced
by the particle on its way through the 
pore are negligible.
But when the Debye length becomes comparable 
to the pore diameter, the 
particle encounters a quite significant
potential barrier while approaching and entering 
the pore, and symmetrically upon exiting the pore.
The main reason is an increasing pressure which acts 
on the particle when it intrudes into the counter
ion cloud of the Debye screening layer.
In case the polarizability of the particle is different
(usually smaller) than that of the ambient fluid,
a second, much smaller contribution to the potential
barrier is due to self-energy effects.
Our numerical treatment of the problem is complemented 
by analytical approximations for sufficiently 
long cylindrical particles and pores, which 
agree very well with the numerics.
\end{abstract}


\maketitle

\section{Introduction}
\label{s1}
\begin{figure}
\begin{center}
    \hspace*{-0.1cm}	
	\epsfxsize=0.5\columnwidth
	\epsfbox{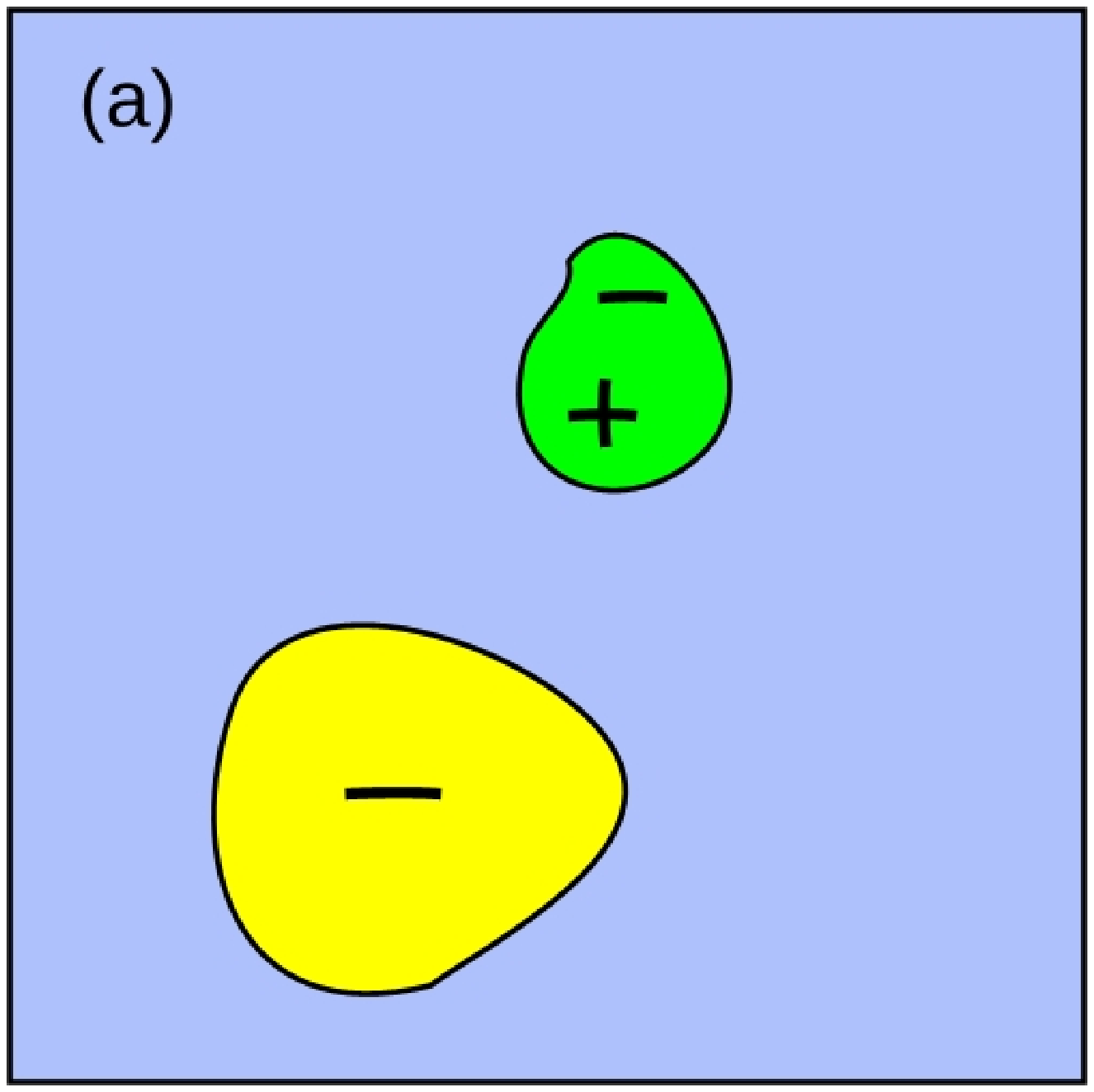}
	\hspace*{-0.2cm}
	\epsfxsize=0.5\columnwidth
	\epsfbox{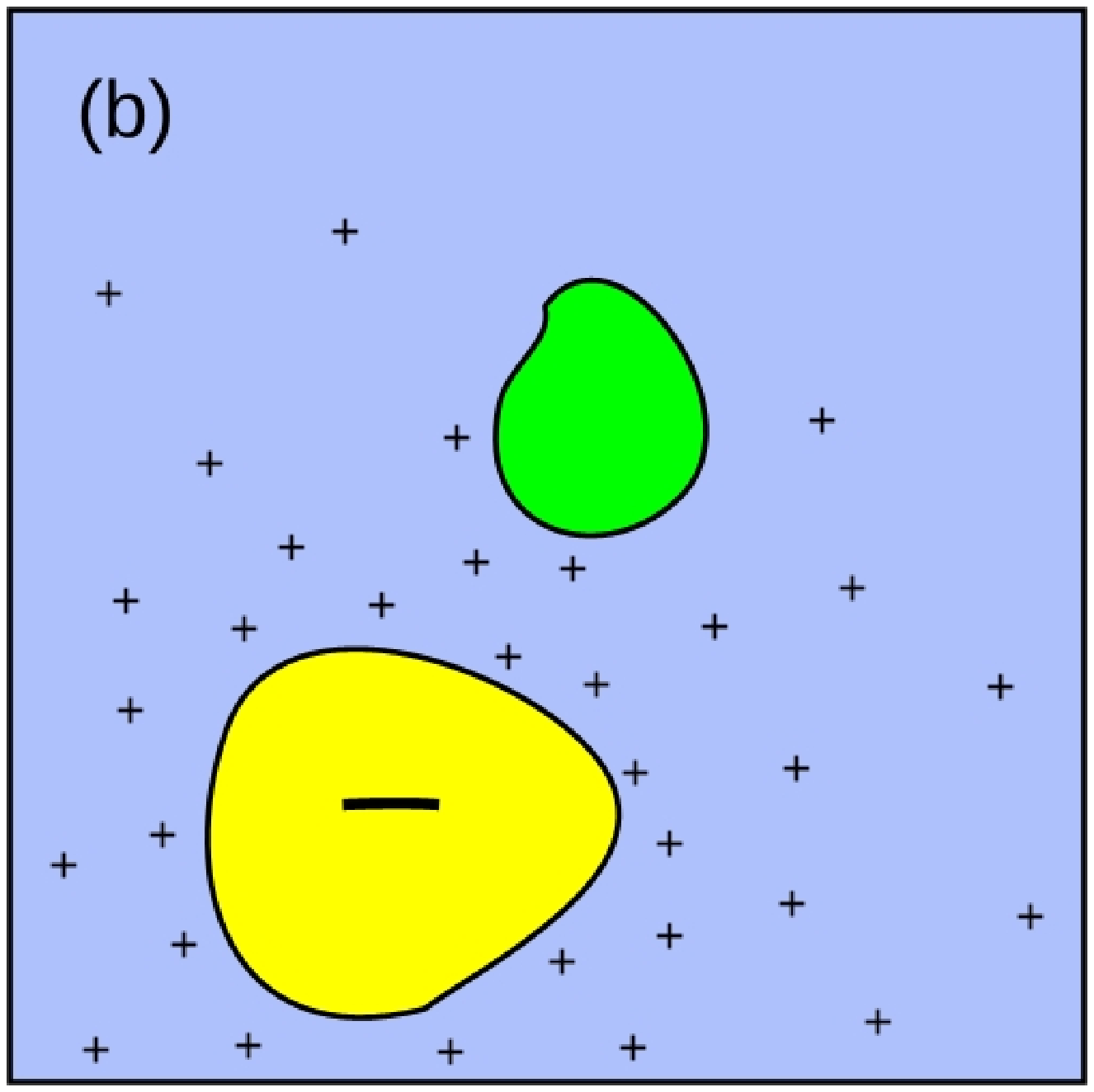}\\[0.3cm]
\end{center}
\caption{\label{fig1}
(Color online) 
Interaction forces between a negatively charged 
and an electrically neutral object.
(a) Dielectric forces: 
The charged object induces a polarization 
(electric dipole) in the nominally neutral 
object, resulting in an attraction between 
the two objects in vacuum.
In the presence of an ambient aqueous solution, whose
polarizability (permittivity) is higher than that of 
the neutral object, the net result
is a repulsion between the two objects.
(b) Counterion pressure:
The negatively charged object attracts positive 
counterions from the ambient electrolyte solution.
The neutral object is repelled by the increased 
pressure within the counterion cloud
(caused by the mutual repulsion of the 
counterions).
Analogous considerations apply for a positively instead 
of a negatively charged object.
}
\end{figure}

How can a charged object 
exert electric forces on 
an uncharged object?
Generally speaking, this may be the case
whenever the nominally neutral object and/or 
its overall neutral environment is actually
composed of positive and 
negative constituents,
which may freely move (e.g. ions in a liquid)
or lead to dielectric polarization effects
(e.g. electrons and protons in a solid) 
in the presence of an electric field,
see Fig. \ref{fig1}.

According to textbook Electrostatics \cite{jac99}, 
any given charge 
in a polarizable environment
gives rise to a so-called ``self-energy'' 
or ``energy of charge'', which is negative
and proportional to the polarizability
of the environment.
As a consequence, a repulsive force between
the given charge and any other extended
object arises in the most common
case that they are both surrounded by
a fluid and the polarizability 
of the object is lower than that of 
the fluid, see Fig. \ref{fig1} (a).

One of the best known examples is an ion (charged object)
in an aqueous solution (dielectric constant of water 
$\epsilon_w\approx 80$)
being repelled by a biological or artificial 
membrane (neutral object) with a typical 
dielectric constant $\epsilon_m$ of $2-5$ 
\cite{par69,jor89,zha07,bon06,kes11}.
Here, we address the
conceptually analogous
case of the potential barrier which an {\em uncharged}
particle encounters upon entering
a charged membrane pore.
The corresponding forces on the particle 
are closely related to the well-known 
phenomenon of dielectrophoresis 
and will henceforth be denoted 
as {\em dielectric forces}.

Additional important effects arise
in the usual case that the ambient fluid
contains positively and negatively charged ions 
\cite{jor89,zha07,bon06,kes11}.
As a consequence, any charged object
now attracts counterions (and repels 
coions) from the ambient fluid
(electric double layer).
The characteristic extension of such a counterion 
``layer'' or ``cloud'' is quantified by the so-called 
Debye length, typically of the order of $1\,$ nm.
At distances beyond a few Debye lengths, the charged 
object is thus essentially screened, i.e. it ``looks''
as if it were uncharged.

On the one hand, the above mentioned dielectric
forces are therefore expected to be notably reduced already 
within the counterion layer and become negligible 
outside it.
On the other hand, additional repulsive
forces are expected when an uncharged 
object enters the electric double layer,
see Fig. \ref{fig1} (b).
The reason is that the prevailing like-charged 
ions repel each other,
resulting in an excess pressure 
within the counterion
cloud which also acts against any
intruding object.
While this effect bears some resemblance
to {\em osmotic pressure} \cite{sad00}, the term 
{\em counterion pressure} seems more 
appropriate to us and will be adopted
from now on.

A detailed exploration of those  
effects is the main subject of our present work.
In particular, we will demonstrate that, under
typical experimental conditions, the counterion
pressure gives rise to considerably larger
energy barriers against the entrance of 
uncharged nanoparticles into charged 
nanopores than the dielectric forces.

\section{Model}
\label{s2}

\begin{figure}
\begin{center}
	\epsfxsize=0.4\columnwidth
	\epsfbox{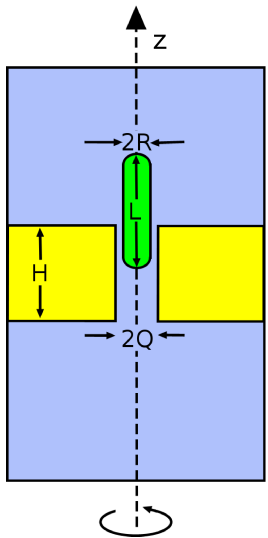} 
\end{center}
\caption{\label{fig2}
(Color online) Schematic illustration of 
the model.
A membrane of thickness $H$ separates two reservoirs 
which are filled with electrolyte solution and  
connected by a cylindrical nanopore of radius $Q$. 
A prolate particle with radius $R$ and length $L$ 
translocates through the pore along the $z$-axis (dashed line).
The complete setup is rotationally symmetric about the $z$-axis,
whose origin ($z=0$) is at the pore center.
}
\end{figure}

A fluid chamber is divided 
into two compartments by a thin biological or solid-state membrane 
(typical thickness $H\approx20$ nm)
with a cylindrical pore (typical radius $Q\approx 5$ nm)
\footnote{For the sake of both numerical 
convenience and a more realistic modeling,
the ``corners'' of the pore in Fig. \ref{fig2}
are in fact slightly rounded ($1\,$nm curvature radius).},
see Fig. \ref{fig1} and Refs. \cite{mel03,kas96,mel01,but07,dek07,sto05,wan08,spi11}.
The fluid chamber contains an electrolyte 
solution with $N_\nu$ different ionic species
and preset bulk concentrations $c_{\nu,0}$
(far away from the membrane).
Typical experimental concentrations $c_{\nu,0}$ range
from about $10$ mM to about $1000$ mM 
\cite{pro03,mel03,kas96,mel01,but07,dek07,sto05,wan08,spi11}.
The membrane is modeled as an insulator with constant 
surface charge density $\sigma$, 
whose quantitative value
depends on the membrane material, pH-value, salt concentrations,
and other factors, and which is screened by counterions 
in the electrolyte solution. 
Under typical experimental conditions the surface charge density
$\sigma$ is of the order of $\sigma  = -50$ mC/m$^2$ 
\cite{beh01,ste04,he11,kir04,and11,hoo09}.

An uncharged, prolate particle of radius $R<Q$ and length $L\ge 2R$ 
is contained in the electrolyte solution.
In our model, the particle is a cylinder with half spheres merged to 
the ends, and hence the particle becomes a sphere if $L=2R$.
Typical values in our examples below will be $R=3$ nm
and $L=6 - 60$ nm.

The quantity of foremost interest is the
net force acting on the particle due to 
the counterion pressure and the dielectric forces
mentioned in Sec. \ref{s1}.

From the viewpoint of the numerical
(and analytical) tractability of the problem, 
two further assumptions are practically unavoidable
\cite{ai10,ai11,zha12,kes11},
see also Sec. \ref{s3e}.
First, we restrict our discussion to the 
axisymmetric case where a particle translocates 
through the pore along the $z$-axis 
(see Fig. \ref{fig2}). 
This will also be justified by our later
finding that the neutral particle is repelled 
by the uniformly charged membrane and pore 
surfaces and hence the energetically most
favorable translocation path 
(e.g. driven by thermal noise)
will be along the pore axis
\cite{he11,che10,ai11,lua11,lua08}.
Second, we restrict ourselves to
steady state (time-independent) 
situations, and we assume, similarly 
as in Refs. \cite{ai11,kes11,lua11,lua08},
that the particle itself
does not to exhibit any notable proper
motion within the relaxation time
of its environment.
In other words, the particle position 
is a model parameter 
rather than a dynamical variable.
The justification is, as usual, the
clear cut time- and length-scale
separation between the nanoparticle 
and the molecular degrees of freedom 
of the ambient fluid \cite{bon06,kes11,lub99}.

\section{General framework}
\label{s3}
\subsection{Basic equations}
\label{s3a}
In this section we summarize the Poisson, Nernst-Planck,
and Stokes equations.
A more detailed discussion is
provided e.g. in Refs. \cite{eis96,cor00,pro03,mas06}.

The electric potential $\psi$ obeys the Poisson equation
\begin{equation}
\label{1}
-\epsilon_0 \nabla \cdot \left[\epsilon(\bx) \nabla \psi(\bx)\right]
= \rho(\bx) + \rho_m(\bx)  \ ,
\end{equation}
where $\epsilon_0$ is the vacuum permittivity and $\epsilon$ the 
dielectric constant (relative permittivity).
While the static charge density $\rho_m$ is associated with 
the fixed membrane surface charges, 
$\rho$ denotes the local charge density due to the mobile ions,
\begin{equation}
\label{3}
\rho(\bx) := 
\begin{cases}
e_0\,\NA \sum\limits_{\nu=1}^{N_\nu} Z_{\nu} c_{\nu}(\bx)\  & 
\text{in the electrolyte solution} \\
0\  &
\text{else}
\end{cases}	
\end{equation}
Here, $e_0=1.602...\times 10^{-19}\,$C is the elementary 
charge and $\NA=6.022...\times 10^{23}$/mol is Avogadro's constant.
Furthermore, $Z_{\nu}$ and $c_{\nu}$ denote
the valence and the molar concentration
of the $\nu^{\mathrm{th}}$ ionic species.

The particle flux  density
$\bJ_\nu$ of the $\nu^{\mathrm{th}}$ ionic species consists of 
three different contributions, arising from 
(i) convection due to the velocity field of the fluid $\bu$, 
(ii) concentration gradients, 
and (iii) the electric field ${\bf E}=-\nabla \psi$, 
and is given by the Nernst-Planck equation
\begin{equation}
\label{4}
	\bJ_\nu(\bx) = c_\nu(\bx) \bu(\bx) - D_\nu\nabla c_\nu(\bx) 
	- \mu_\nu c_\nu(\bx) \nabla \psi(\bx) \ .
\end{equation}
The diffusion coefficient $D_\nu$ is related to the mobility 
$\mu_\nu$ via $\mu_\nu = Z_\nu e_0\, D_\nu /\kB T$, where 
$\kB$ denotes Boltzmann's constant and $T$ the temperature.
Assuming a steady state,
particle number conservation implies
\begin{equation}
\label{5}
\nabla \cdot \bJ_\nu(\bx) = 0 \ .
\end{equation}

The velocity field $\bu$ and the pressure $p$ of the electrolyte solution
are governed by the Navier-Stokes equation. 
Since in our present case  the Reynolds number is very low, 
the non-linear terms in this equation can be safely neglected.
Focusing on steady states and neglecting the very small effect of 
gravity, it is thus sufficient to solve the simpler Stokes equation
\begin{equation}
\label{6}
\eta \, \Delta \bu(\bx) - \nabla p(\bx) -\rho(\bx) \nabla \psi(\bx) = {\bf 0} \ ,
\end{equation}
with $\eta$ the viscosity of the fluid. 
Analogously to Eq. (\ref{5}), 
assuming an incompressible fluid and
focusing on steady state solutions, mass conservation implies
\begin{equation}
\label{7}
\nabla \cdot \bu(\bx) = 0  \ .
\end{equation}

\subsection{Parameters and boundary conditions}
\label{s3b}

The relative permittivity is assumed to be of the form
\begin{equation}
\label{8}
\epsilon (\bx) = 
\begin{cases}
\epsilon_w \  &\text{in the electrolyte solution}\\
\epsilon_m \  &\text{in the membrane}\\
\epsilon_p \  &\text{in the particle}
\end{cases}
\end{equation}
While $\epsilon_w$ is usually close to the value $\approx 80$ for water at room temperature,
$\epsilon_m$ and $\epsilon_p$ are material dependent.
A typical value for biological matter is $\epsilon = 2$ and for 
solid-state membranes (e.g. SiO$_2$, SiN, Si$_3$N$_4$) $\epsilon = 5$  
\cite{par69,jor89,zha05,bon06,kes11,sad00,zha07,che08,dor09,das12}.
Typical particle permittivities $\epsilon_p$ are in the
same range and actually turn out to play a 
very minor role.
Thus, we will usually adopt the following choices: 
$\epsilon_w = 80$ and $\epsilon_m=\epsilon_p = 5$.

With respect to the static (fixed) surface charges we can, as
detailed in \cite{jac99,kes11}, formally set $\rho_m =0$ 
in Eq. (\ref{1}) and instead work with the boundary conditions 
\begin{eqnarray}
	\label{9}
	\psi(\bx) \ \text{continuous} \ , \\
	\label{10a}
	\left(\epsilon_w\nabla \psi_w(\bx) - \epsilon_{m}\nabla \psi_{m}(\bx)
	\right) \cdot {\bf n}(\bx) = - \sigma /\epsilon_0 \ , \\
	\label{10b}
	\left(\epsilon_w\nabla \psi_w(\bx) - \epsilon_{p}\nabla \psi_{p}(\bx)
	\right) \cdot {\bf n}(\bx) = 0 \ ,
\end{eqnarray}  
where $\psi_w$, $\psi_{m}$, and $\psi_{p}$ denote the electric
potential at the respective
side of the interface (see also Eq. (\ref{8}))
and where ${\bf n}$ is the normal vector pointing from the membrane 
and the particle, respectively, into the electrolyte solution.
The remaining boundary conditions for the Poisson equation (\ref{1})
are chosen as $\psi(\bx) =0$ at the top and bottom walls of the
cylindrical fluid chamber in Fig. \ref{fig2} and
$\nabla \psi(\bx) \cdot {\bf n}(\bx) = 0$ at the 
cylindrical side wall.

Turning to the Nernst-Planck equation (\ref{4}), the 
concentrations $c_\nu$ of the different ionic species are 
required to assume their bulk values $c_{\nu,0}$ at the top and
bottom walls of the fluid chamber.
On the membrane and particle walls as well as on
the cylindrical side wall of the fluid chamber,
we impose insulation (or reflecting) boundary 
conditions $\bJ_\nu (\bx) \cdot {\bf n}(\bx)=0$.
In the following, we restrict our discussion to 
so-called symmetric $Z:Z$ 
electrolytes, i.e. $N_\nu=2$, $Z_1=-Z_2=:Z$, and
$c_{1,0}= c_{2,0}=: c_0$.
Moreover, in our numerical examples below we will 
focus on the simplest case $Z=1$. 
We further adopt 
a typical temperature $T=300$ K and 
typical diffusion coefficients 
$D_1=D_2=:D=2\cdot10^{-9}$ m$^2$/s \cite{pro03}. 
Accordingly, the Debye length, quantifying
the extension of the electric double layer
(see also Secs. \ref{s1} and \ref{s4a})
takes the form \cite{pro03}
\begin{equation}
\lambda_D =\sqrt{\frac{\epsilon_0\epsilon_w\kB T}{2 e_0^2\NA c_0}}
\simeq \sqrt{1000\,\mbox{mM}/c_0}\ 0.305\, \mbox{nm} \ .
\label{deb}
\end{equation}

With respect to the Stokes equation (\ref{6}),
our first assumption is that
the usual no-slip boundary conditions $\bu = {\bf 0}$
are satisfied on the membrane and particle walls as well as on
the cylindrical side wall of the fluid chamber
(see Fig. \ref{fig2}).
Concerning the top and bottom walls of the 
fluid chamber, 
we require that the pressure $p$ 
approaches some preset ``bulk value'' $p_0$,
\begin{eqnarray}
	\label{11}
	p(\bx) &=& p_0 \ .
\end{eqnarray}  
Since only the gradient of $p$ matters (see Eq. (\ref{6})),
we can and will set 
\begin{equation}
p_0 =0
\label{11a}
\end{equation} 
without loss of generality.
Moreover, we require that the normal component of the 
hydrodynamic stress (see e.g. \cite{mas06}
for a more detailed discussion) vanishes at 
the top and bottom walls of the fluid chamber,
\begin{eqnarray}
	\label{12}
	\left[ -p(\bx) +
	\eta\left(
		\nabla \bu(\bx) + [\nabla \bu(\bx)]^T
	\right)\right]{\bf n}(\bx) &=& {\bf 0}  \ ,
\end{eqnarray}
where $\nabla \bu$ denotes the matrix with elements
$\partial u_i/\partial x_j$ and $[\nabla \bu]^T$ the
transposed matrix.
The boundary conditions (\ref{11})-(\ref{12}) are well-known 
to be numerically stable provided the boundaries are 
sufficiently far away from the nanopore 
\cite{pan05,com12}.
Quantitatively, we found that for not too large 
particle lengths ($L\lesssim H$), and not too low
concentrations ($c_0\gtrsim1\,$mM),
finite-size effects 
become negligible for fluid chambers  
(see Fig. \ref{fig2}) beyond a radius 
of about 40 nm and a height of about 100nm. 
For larger particle lengths or lower concentrations,
the size of the fluid chamber was increased, so that
finite size effects were again negligible.
 
For all examples, the viscosity of water at room temperature
takes its standard value $\eta = 10^{-3}\,$Pa$\cdot $s
and the values of the remaining 
parameters will be specified later.

\subsection{Forces on the particle}
\label{s3c}

The force ${\bf F}$ acting on the particle in Fig. \ref{fig2}
can be decomposed into two contributions \cite{mas06}, 
one arising from the interaction with the
electric field ${\bf E} = -\nabla \psi$,
and the other from the hydrodynamic
interaction with the surrounding
electrolyte solution,
\begin{equation}
	\label{15}
	{\bf F} = {\bf F}_e + {\bf F}_h \ .
\end{equation}
The electrostatic force ${\bf F}_e$ can be
calculated quite generally 
\cite{jac99,sad00}
by integrating the
Maxwell stress tensor over the particle surface
${\mathcal O}(P)$,
\begin{equation}
	\label{16}
	{\bf F}_e = \epsilon_w\epsilon_0
	\oint\limits_{{\mathcal O}(P)} 
	\left(\bE(\bx)\, [\bE(\bx)]^T-\frac{1}{2} |\bE (\bx)|^2
	\right){\bf n}(\bx) dS \ .
\end{equation}
Similarly, the hydrodynamic force ${\bf F}_h$ follows by
integrating the hydrodynamic stress tensor
(also called pressure tensor) over the particle surface,
\begin{equation}
	\label{17}
	{\bf F}_h = 
	\oint\limits_{{\mathcal O}(P)} 
	\left(
	-p({\bf x}) + \eta\left(
		\nabla \bu(\bx) + [\nabla \bu(\bx)]^T\right)
	\right){\bf n}(\bx) dS \ .
\end{equation}

If $\epsilon_p=\epsilon_w$, the net electric 
force ${\bf F}_e$ on the particle must vanish.
Physically, this follows by observing that for an
overall constant $\epsilon$, the electric field
is proportional to the field for $\epsilon=1$, 
i.e. for a non-polarizable particle in vacuum.
Since there 
are no dielectric forces in the latter case, the same applies to the former case.
(${\bf F}_e$ in (\ref{16}) must be zero in both cases).
Formally, the same follows from Poisson's equation (\ref{1}),
whose right hand side vanishes within the particle region,
the boundary conditions (\ref{9}), (\ref{10b}),
which imply a smooth behavior of $\bE$ across the
particle boundary for $\epsilon_p=\epsilon_w$,
and by employing the Gauss theorem in (\ref{16}).
For a neutral particle with arbitrary $\epsilon_p$ 
it thus seems justified to henceforth {\em identify
${\bf F}_e$ with the dielectric forces}
due to polarization or self-energy effects 
from Fig. \ref{fig1}a and Sec. \ref{s1}.

\subsection{Simplifications at thermal equilibrium}
\label{s3d}
Since we are considering an isolated system 
(no external forces are acting) and we focus on 
steady state solutions, this steady state must
be tantamount to thermal equilibrium.
Hence all macroscopic fluxes in the system 
must vanish, i.e.
\begin{equation}
\label{18}
\bu(\bx) = {\bf 0} \ \text{and}\ \bJ_\nu(\bx) = {\bf 0} \ .
\end{equation} 
As a consequence, all boundary conditions 
involving $\bJ_\nu$ and $\bu$ are automatically
fulfilled.
Furthermore, the Nernst-Planck equation (\ref{4}) 
is formally solved by the Boltzmann-distribution
\begin{equation}
	\label{19}
	c_\nu(\bx) = c_{\nu,0} \exp\left(
		-\frac{Z_\nu e_0\psi(\bx)}{\kB T} \right) \ ,
\end{equation} 
while Eq. (\ref{5}) is  trivially satisfied.
Likewise, the Stokes equation (\ref{6}) with boundary 
conditions (\ref{11}) and (\ref{11a}) can be formally
intergated, yielding for the pressure the result
\begin{equation}
	\label{20b}
	p(\bx) = 2\kB T \NA c_0 \left[\cosh\left(
	\frac{Ze_0\psi(\bx)}{\kB T}  
	\right)-1\right] \ .
\end{equation}

The only remaining equation is thus Poisson's equation (\ref{1}).
Concerning the two charge density terms which appear on the right 
hand side of this equation (\ref{1}) we observe that:
(i) Exploiting (\ref{19}), the ion charge density (\ref{3}) in a
$Z$:$Z$ electrolyte solution (see above Eq. (\ref{deb}))
can be rewritten as 
\begin{equation}
	\label{20a}
	\rho({\bf x})
	= - 2Z e_0\NA c_0
	\sinh\left(
	\frac{Ze_0\psi({\bf x})}{\kB T}
	\right)
    \ .
\end{equation} 
(ii) The fixed membrane surface charges $\rho_m$ are effectively
accounted for by the boundary conditions (\ref{9})-(\ref{10b}).
By combining (\ref{1}) and (\ref{20a}), we 
are thus left with the so-called Possion-Boltzmann 
equation for the electric potential $\psi$
\begin{equation}
	\label{20a1}
\epsilon_0 \nabla \cdot \left[\epsilon(\bx) \nabla \psi(\bx)\right]
= 2Z e_0\NA c_0
	\sinh\left(\frac{Ze_0\psi({\bf x})}{\kB T}\right)
    \ .
\end{equation} 
Once this equation with (\ref{8})-(\ref{10b}) is solved, the
concentration, pressure, and charge density fields immediately 
follow from (\ref{19})-(\ref{20a}).

Turning to the forces, we first remark that
${\bf F}_h$ from (\ref{17}) simplifies to the
familiar pressure integral
\begin{equation}
	\label{17a}
	{\bf F}_h = -
	\oint\limits_{{\mathcal O}(P)} 
	p({\bf x})\, {\bf n}(\bx) dS \ .
\end{equation}
In other words, ${\bf F}_h$ {\em quantifies the 
counterion pressure effects} from
Fig. \ref{fig1}b and Sec. \ref{s1}.

Furthermore, for symmetry reasons 
the force ${\bf F}$ resulting from (\ref{15}), (\ref{16}), (\ref{17a}) 
will be parallel to the pore axis.
Henceforth, this force component is
denoted by $F(z)$ for any given 
position $z$ of the particle center in Fig. \ref{fig2}.
Finally, the corresponding potential energy 
$U(z)$, from which $F(z)$ derives, follows as
\begin{equation}
	\label{36}
	U(z) = -\int\limits_{z_0}^{z} dz'\;F(z') \ ,
\end{equation}
where $z_0$ denotes the particle position when it
touches the bottom wall in Fig. \ref{fig2}.
Symmetry reasons further imply that
\begin{equation}
F(-z)=-F(z) \ .
\label{ss1}
\end{equation}
With (\ref{36}) we thus can conclude that
\begin{equation}
U(-z)=U(z) \ .
\label{ss2}
\end{equation}

\subsection{Numerical method}
\label{s3e}
In spite of the above mentioned simplifications at thermal
equilibrium, the remaining Poisson-Boltzmann equation (\ref{20a1}),
complemented by (\ref{8})-(\ref{10b}),
and the final surface integrations in (\ref{16}), (\ref{17a})
can only be tackled analytically
in a few special cases and within certain 
approximations, see e.g.  Sec. \ref{s4}.
In all other cases, only numerical solutions
are possible.
It, however, turns out that the numerical 
treatment of the fully three-dimensional problem 
is still very demanding, even on modern computers, 
if a satisfactory numerical accuracy 
is required.
Consequently, similar previous studies are restricted to 
one-dimensional \cite{dor09}, two-dimensional \cite{ai11}, 
or axisymmetric problems
\cite{das03,che11,zha12,bow96,get13} problems.
In the latter case, which is also at the focus 
of our present work,
an effectively two-dimensional problem is
readily recovered when going over to 
cylindrical coordinates (see e.g. \cite{mas06} 
for the explicit expressions).
Our numerical results presented below were 
obtained using the commercial COMSOL 4.3a 
Multiphysics package of coupled
partial differential equation 
solvers, exploiting finite element 
methods \cite{com12}.

\section{Analytical Approximations}
\label{s4}
For a very long particle ($L\gg H$ in Fig. \ref{fig2}),
which is ``fully threaded through the pore'' so that
both ends stick far out at
either side of the pore,
we are dealing with an almost translation invariant situation 
and the net force on the particle will be practically zero.

Likewise, for a very ``long'' pore ($H\gg Q$ in Fig. \ref{fig2})
and a comparatively ``short'' particle ($L\ll H$)
with both ends ``far inside'' the pore,
an almost translation invariant net force on the particle
is expected (the pore ends hardly matter).
Furthermore, for symmetry reasons the forces acting
onto either end of the particle will almost cancel 
each other, i.e. a close 
to zero net force is expected.

The main focus of the present section is on
the following ``mixture'' of the above two cases:
We consider a
very ``long'' pore ($H\gg Q$ in Fig. \ref{fig2})
in combination with a sufficiently ``long'' 
particle ($L\gg Q$),
so that one of its ends is ``far inside'' the pore
and the other end ``far outside'' the pore.
Again, one thus expects an almost translation
invariant net force on the particle, but now 
there is no symmetry argument that this
constant force should be almost zero.
Rather
one expects that the force will actually be (almost) 
maximal (in modulus).
In the following, our main goal is to 
analytically approximate this ``maximum force'',
henceforth denoted as $F^m$.

Closer inspection of the surface integrals in Eqs. (\ref{16})
and (\ref{17a}) shows that, as expected from our
above considerations, the main contributions to $F^m$ are
generated in the vicinity of the particle's end far 
inside the pore.
Since no analytical (exact or approximate) solutions 
of the Poisson-Boltzmann equation (\ref{20a1}) 
seem to exist for such a case, we cannot evaluate 
the surface integrals (\ref{16},\ref{17a}) directly.
We therefore adopt the following, alternative approach:
We assume that the particle is moved an infinitesimal 
distance $\Delta z$ into the pore (along the cylinder axis).   
Accordingly, the free energy of the system will change by 
an amount $\Delta G$. The force $F^m$ required to hold the 
particle fixed at the initial position is thus given by
\begin{equation}
	\label{20c}
	F^m = -\Delta G/\Delta z 
\ .
\end{equation}

In the framework of the Poisson-Boltzmann 
equation (\ref{20a1}) the 
free energy of the system can be expressed in several
equivalent forms \cite{sha90,rei91}. 
The most convenient form for our purpose is
\begin{equation}
	\label{20d}	
	G = \int\limits_{V}d\bx\,\left[
		\frac{\rho_m(\bx)-\rho(\bx) } {2}\, \psi(\bx)
		-p(\bx)
		\right]  \ ,
\end{equation}
where the integration domain $V$ is
the entire fluid chamber.
Since $H,\, L \gg Q$, we can approximate the free energy difference as
\begin{equation}
	\label{20e}
	\Delta G \approx (g_2 - g_1)\Delta z \ ,
\end{equation}
and hence $F^m$ from (\ref{20c}) as
\begin{equation}
	\label{23}
	F^m \approx g_1-g_2 \ ,
\end{equation}
where $g_1$ is the free energy per unit length for an
infinitely long, empty pore, and 
$g_2$ is the corresponding free energy per unit length
for an infinitely long particle 
in an infinitely long pore.
For both situations the potentials $\psi_{1,2}$ and the charge densities 
$\rho_{1,2}$ become independent of $z$.
We may thus 
adopt cylinder coordinates with
$r := (\left|\bx\right|^2-z^2)^{1/2}$ 
and rewrite the
Poisson-Boltzmann equation (\ref{20a1}) as
\begin{equation}
	\label{21}
	\frac{1}{r}\dr \left(r\dr \psi_{1,2}(r)\right) 
	= -\frac{2Ze_0\NA c_0}{\epsilon_0\epsilon_w} 
	\sinh\left(
	\frac{Ze_0\psi_{1,2}(r)}{\kB T}
	\right)
    \ ,
\end{equation} 
complemented by the conditions
$\psi_{1,2}(r)=0$ for $r>Q$ (overall charge neutrality of
pore surface and counterions, cf. Eq. (\ref{1}) and
convention $\psi_{1,2}(r\to\infty)=0$),
$\psi_{1,2}'(r=Q) = \sigma/\epsilon_0\epsilon_w$ 
(cf. Eq. (\ref{10a})),
$\psi_1'(r=0)=0$ (regularity at the pore center),
$\psi_2'(r=R)=0$ (cf. Eq. (\ref{10b})), and
$\psi_2'(r)=0$ for $r<R$ (cf. Eq. (\ref{1})).

While equation (\ref{21}) still cannot be solved 
analytically in full generality, we will focus on
approximations for two limiting cases in the following
two subsections.

Once the latter problem is solved,
we can exploit that $\rho(\bx)=0$ outside the electrolyte 
solution and that $\rho_m(r) = \sigma\delta(r-Q)$ to
calculate the free energy per unit length according 
to Eq. (\ref{20d}) as
\begin{equation}
	\label{21a}	
	g_i = \pi\sigma Q\psi_i(Q) - \int\limits_{l_i}^Q dr\, 2\pi r\left[
		\frac{\rho_i(r)\psi_i(r)}{2}
		+p_i(r)
		\right]  
\end{equation}
with $l_1=0$ and $l_2=R$.
Finally, $F^m$ follows according to (\ref{23}).

\subsection{High concentration or low surface charge}
\label{s4a} 
We first focus on the so-called Debye-H\"uckel limit $Ze_0\left|\psi_i(r)\right|/\kB T \lesssim 1$
throughout the nanopore, which is tantamount to
low surface charge densities $\sigma$
and/or high bulk concentrations $c_0$.
The corresponding approximation for the
the maximum force $F^m$ from (\ref{23})
is henceforth denoted as $F_l^m$.
Referring to the Appendix I for the detailed 
calculation, the final result is
\begin{eqnarray}
F_l^m & = & \frac{\pi \sigma^2 Q}{\epsilon_0\epsilon_w\kappa} 
\left(
\frac{I_0(\kappa r)}{I_1(\kappa Q)} \right.
\nonumber
\\
& - & \left.
\frac{I_0(\kappa r)K_1(\kappa R)+K_0(\kappa r)I_1(\kappa R)}
{K_1(\kappa R)I_1(\kappa Q)-I_1(\kappa R)K_1(\kappa Q)}
\right)
\label{a1}
\end{eqnarray}
where $\kappa := \lambda_D^{-1}$ is
the inverse Debye length from (\ref{deb}) and
$I_k$ ($K_k$) is the modified Bessel function 
of the first (second) kind and order $k$.

Likewise, for the electric field and the
pressure in the empty pore (index 1)
the following approximations are derived 
in Appendix I:
\begin{eqnarray}
	 \psi_1(r) & = & \frac{\sigma}{\epsilon_0\epsilon_w\kappa} 
	\frac{I_0(\kappa r)}{I_1(\kappa Q)} 
\label{a2}
\\
	p_1(r) & = & \frac{\NA c_0}{\kB T} 
\left(\frac{Ze_0 \sigma}{\epsilon_0\epsilon_w\kappa} 
	\frac{I_0(\kappa r)}{I_1(\kappa Q)}  
	\right)^2 \ .
\label{a3}
\end{eqnarray}

\subsection{Low concentration or high surface charge}
\label{s4b}
Next we turn to the case 
$Ze_0\left|\psi_i(r)\right|/\kB T \gtrsim 1$,
i.e. high surface charge densities $\sigma$
and/or low concentrations $c_0$.
Hence, we follow Philip and Wooding \cite{phi70}
and exploit the approximation
\begin{equation}
	\label{29}
	\sinh\left(
	\frac{Ze_0\psi_i(r)}{\kB T}
	\right)
	\approx
	\frac{\sign(\psi_i(r))}{2}\exp\left(
	\frac{Ze_0\left|\psi_i(r)\right|}{\kB T} 
	\right)\ .
\end{equation}
Making use of $\sign(\psi_i(r))=\sign(\sigma)$ 
and hence $\left|\psi_i(r)\right| = \sign(\sigma )\psi_i(r)$,
the first solution ($\psi_1$) of Eq. 
(\ref{21}) with the boundary 
conditions discussed below (\ref{21}) is
\begin{align}
\label{30}
	\psi_1(r) = 
		-\sign(\sigma )U_0\left(\right.
		&\ln\left[
			\frac{c_0Z e_0 \NA a_1}{a_2}
		\right] 
		 \\ \nonumber 
		 + 2 & \ln\left[
			1-\frac{Z e_0 \left| \sigma  \right| r^2}{a_1}
		\right]
	\left.\right)	 \ ,
\end{align} 
where we have defined
\begin{align}
	\label{32}
	U_0 \ &:=\  \frac{\kB T}{Ze_0} \ ,\\
	\label{33}
	a_1 \ &:= \ Q\left(4\epsilon_0\epsilon_w\kB T 
		+ Ze_0\left|\sigma \right|Q \right) \ ,\\
	\label{34}	
	a_2 \ &:= \  8\epsilon_0\epsilon_w\kB T\left|\sigma \right| \  .
\end{align}
The somewhat more lengthy expressions for $\psi_2(r)$ 
are provided in Appendix II. 

Given $\psi_i(r)$,
the pressure and the charge density follow 
from  (\ref{20b},\ref{20a}).
With (\ref{29}) and an analogous 
approximation for $\cosh(\cdot)$, 
they take the form
\begin{eqnarray}
	\label{34i}
	\rho_i(r) &=& -\sign(\sigma)Z e_0 \NA c_0
\exp\left\{\frac{\sign(\sigma)\psi_i(r)}{U_0}\right\} \ ,\\
	\label{31}
	p_i(r) &=& \kB T\NA c_0\exp\left\{\frac{\sign(\sigma)\psi_i(r)}{U_0}\right\} \ .
\end{eqnarray}

Finally, by exploiting the above results to 
evaluate the integral (\ref{21a}) we arrive
at our final approximation for the maximum 
force $F^m$ from (\ref{23}), henceforth denoted 
as $F_g^m$. An interesting property of this
approximation $F_g^m$ is demonstrated in
Appendix III, namely that it is independent 
of the bulk concentration $c_0$.

\section{Numerical results}
\label{s5}

\subsection{The empty pore}
\label{s5a}
As a first example we consider the case of an empty pore
(Fig. \ref{fig2} without particle). 
Assuming a typical membrane surface charge density of 
$\sigma  = -50$ mC/m$^2$  
and a relatively low
bulk concentration of $c_0 =1$ mM
(cf. Sec. \ref{s2}),
we have numerically solved the 
Poisson-Boltzmann equation (\ref{20a1})
as detailed in Sec. \ref{s3e}.

\begin{figure}[h]
    \hspace*{-0.1cm}	
	\epsfxsize=0.49\columnwidth
	\epsfbox{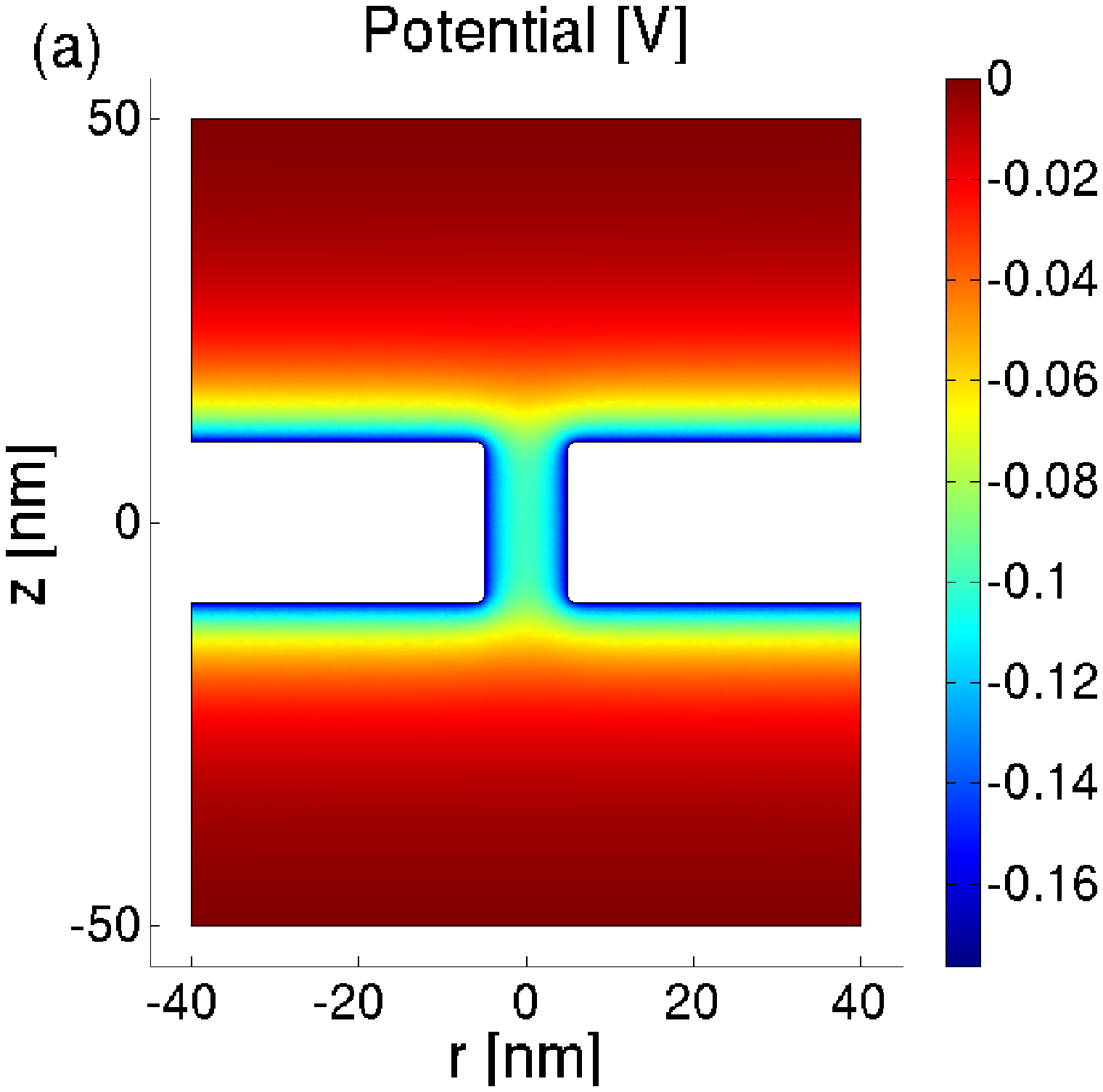}
	\hspace*{-0.1cm}
	\epsfxsize=0.49\columnwidth
	\epsfbox{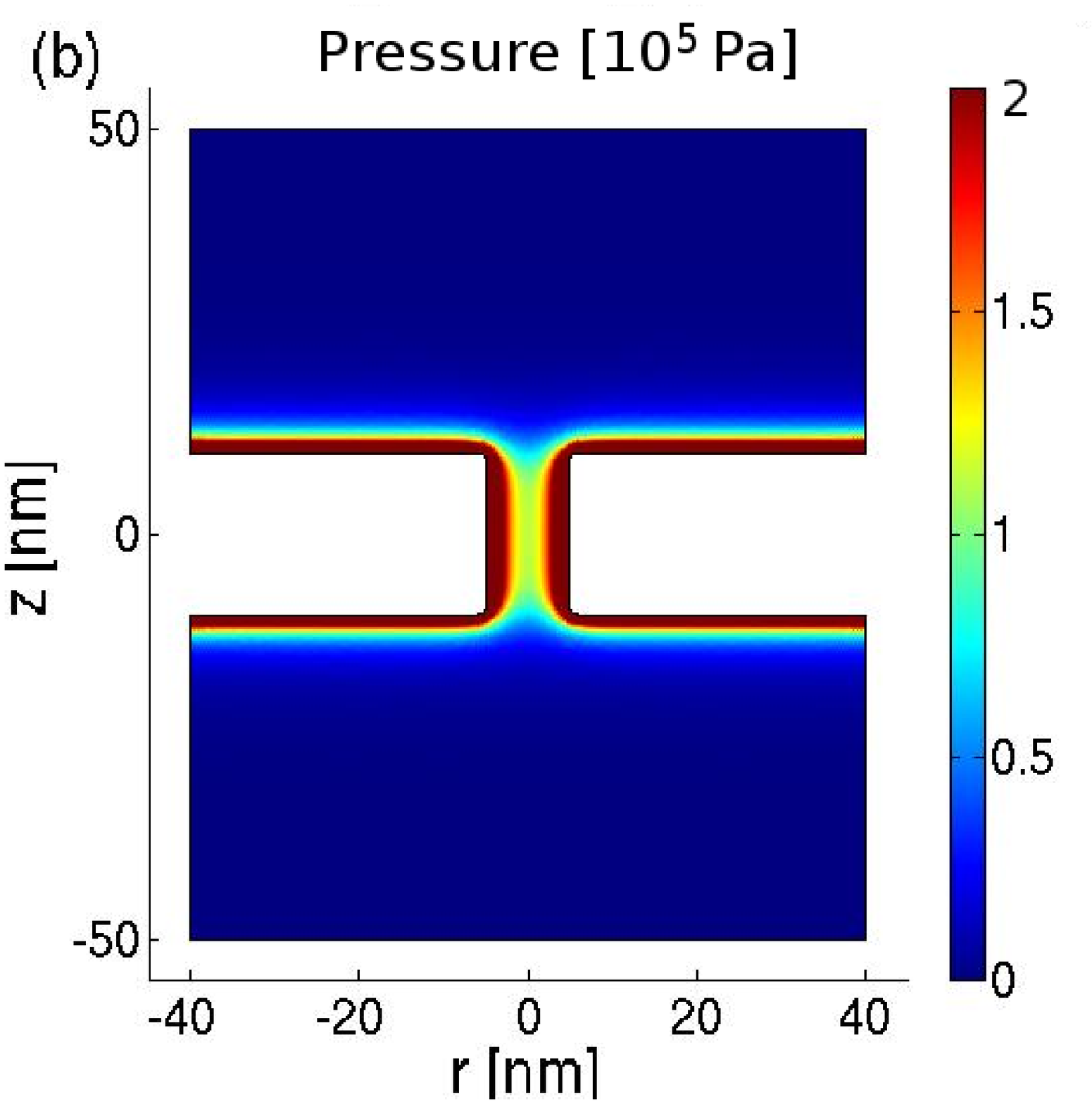}
\\[0.3cm]
\hspace*{-0.5cm}
	\epsfxsize=1.0\columnwidth
	\epsfbox{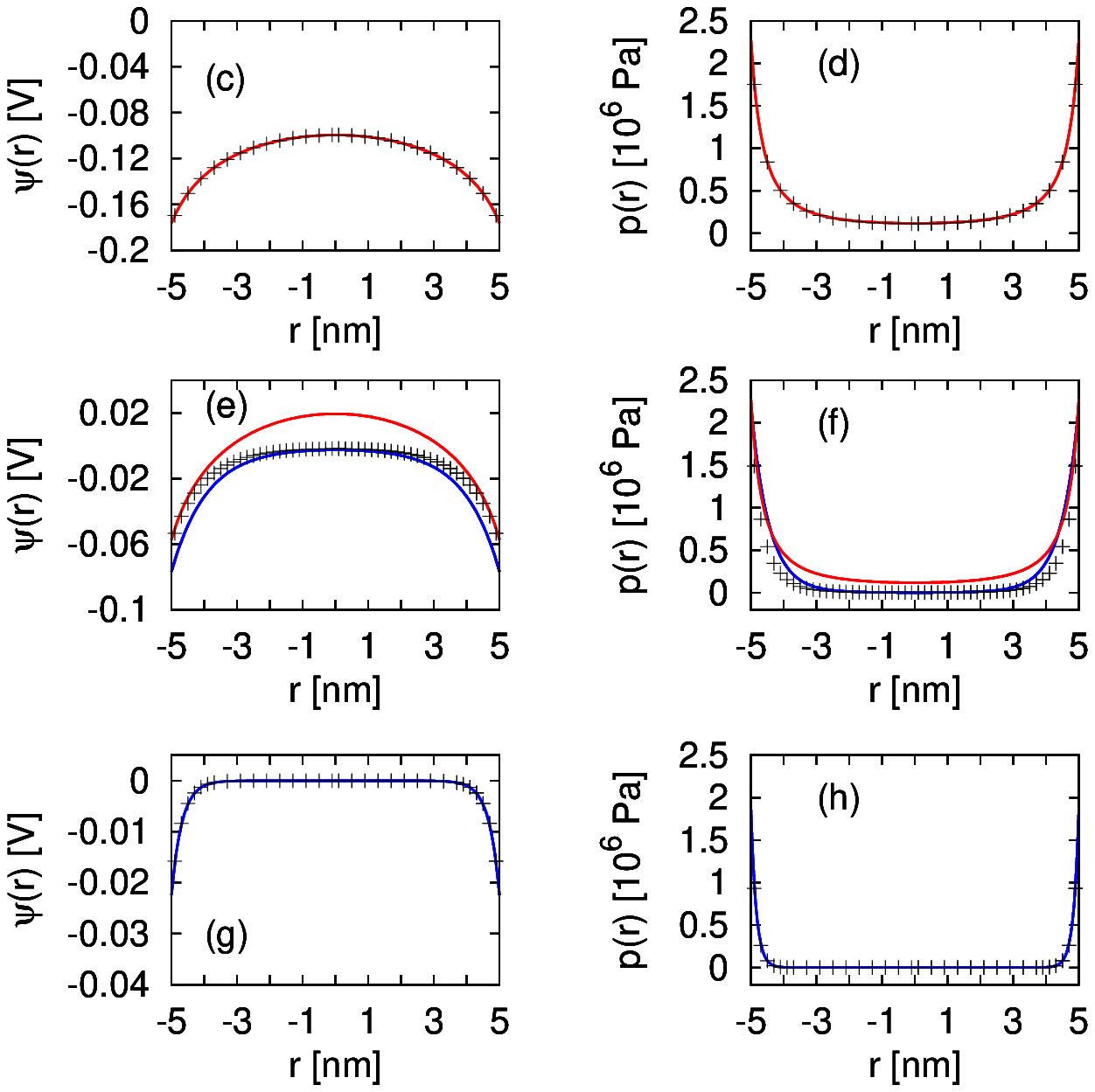}
\vspace*{-0.3cm}
\caption{\label{fig3}
(Color online)
(a) Electric potential $\psi(r,z)$ 
by numerically solving the 
Poisson-Boltzmann equation (\ref{20a1})
for an empty pore (cf. Fig. \ref{fig2})
with radius $Q=5\,$ nm, membrane thickness $H=20\,$ nm,
surface charge density  $\sigma  = -50$ mC/m$^2$,
bulk concentration $c_0 =1$ mM, and 
dielectric constants 
$\epsilon_w = 80$, $\epsilon_m=\epsilon_p = 5$
(cf. Eqs. (\ref{8}-\ref{10b})).
(b) Corresponding pressure 
from (\ref{20b}).
All pressures beyond $2\cdot 10^5$ Pa are displayed in 
deep red (see also (d)).
(c) and (d):
Radial dependence of the same fields for $z=0$
(i.e. in the middle of the pore),
obtained numerically (symbols) and
by means of the analytical approximations
(\ref{a2}) and (\ref{a3}) for low 
concentrations (red lines).
The units in (d) are now $10^6\,$Pa.
(e) and (f):
Same, but for 
$c_0 =100$ mM.
In addition, the analytical approximations
(\ref{30}) and (\ref{31}) for high 
concentrations are shown as blue 
(lower) lines.
(g) and (h):
Same, but for $c_0 =1000$ mM.
}
\end{figure}

Fig. \ref{fig3} illustrates the 
results for the electric potential
$\psi(\bx)$ and for the corresponding 
pressure $p(\bx)$ from (\ref{20b}).
Most remarkably, the pressure within the pore 
increases quite notably beyond the bulk value 
$p_0=0\,$Pa from Eq. (\ref{11a}).
In fact, the counterion pressure near 
the membrane surface typically may 
become as large as 20 bar 
according to Fig. \ref{fig3} (d), (f), (h).

For a cross-section through the $z=0$ plane
(i.e. through the pore center),
Figs. \ref{fig3} (c) and (d) provide a more detailed
picture of the numerically obtained fields 
for $c_0=1$ mM together with the analytical approximations 
(\ref{30}) and (\ref{31}) for low concentrations.

We found a comparably good agreement even
for a tenfold increased bulk concentration, i.e. for
$c_0 =10$ mM (not shown).
Significant deviations arise upon another tenfold 
increased concentration, i.e. for $c_0 =100$ mM,
see red lines in Fig. \ref{fig3} (e) and (f).
Finally, for $c_0=1000$ mM, the analytical approximation
for low concentrations 
is far off the numerics from  Fig. \ref{fig3} (g) and (h).

Likewise, the approximations (\ref{a2}) and (\ref{a3})
for high bulk concentrations reproduce the 
numerical solution almost perfectly for $c_0 =1000$ mM 
(blue lines in Fig. \ref{fig3} (g) and (h)),
develop notable deviations for $c_0 =100$ mM 
(Fig. \ref{fig3} (e) and (f)), 
and completely fail for $c_0 = 10$ mM
and $c_0 =1$ mM.

\subsection{Force on the particle}
\label{s5b}
What happens when an uncharged particle approaches 
and enters the previously considered empty pore? 
On the one hand, the electric potential in 
Fig. \ref{fig3}(a) induces a polarization of the
particle, which, due to the higher polarizability 
of the ambient liquid, results in a net repulsion,
see Fig. \ref{fig1}(a).
On the other hand, the counter ion pressure 
in Fig. \ref{fig3}(b) generates yet another 
repulsive force contribution, see 
Fig. \ref{fig1}(b).
Strictly speaking, there will also be back-reactions
of the approaching particle on the ``unperturbed'' 
electric and pressure fields of the empty pore
from Fig. \ref{fig3}, but these are next-to-leading 
order effects which may be neglected for our present
purpose of a basic understanding of the main mechanisms.

Similarly as at the beginning of Sec. \ref{s4}, both
these forces are expected to develop certain 
``plateau'' regions for sufficiently long pores 
or particles. Concerning a more quantitative understanding, 
especially with respect to the relative importance of the 
two forces, integrating the different contributions 
over the entire particle volume or surface 
(essentially like in (\ref{16}) and (\ref{17a})) 
is unavoidable and goes beyond the realm of
simple intuitive arguments.

\begin{figure}[ht]
\hspace*{-0.5cm}
	\epsfxsize=1.08\columnwidth
	\epsfbox{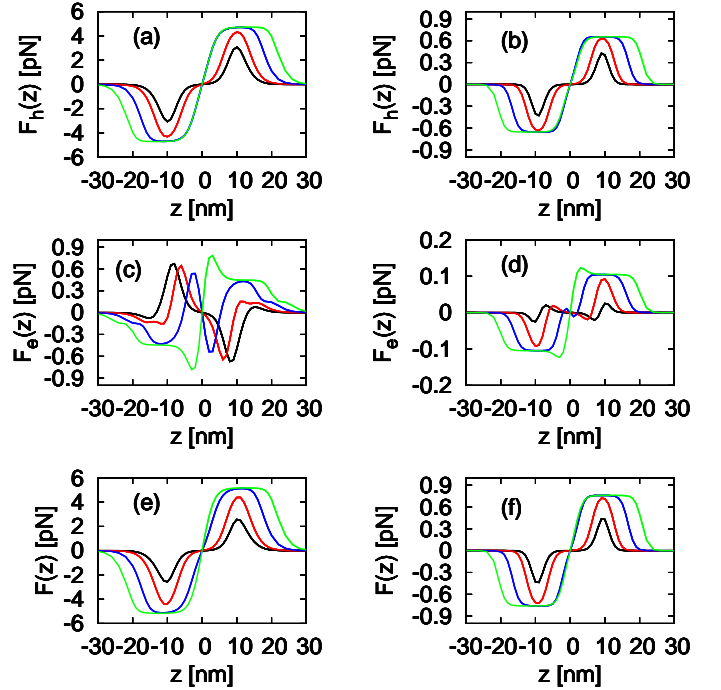}
\caption{\label{fig4}
(Color online)
Numerically obtained results (in units of pN)
versus z (in units of nm)
for the $z$-components $F_e(z)$, $F_h(z)$, and
$F(z)$ of the dielectric force (\ref{16}),
the counterion pressure force (\ref{17a}),
and the total force (\ref{15}),
respectively, when the particle center in Fig. \ref{fig2}
finds itself at an arbitrary but fixed 
position $z$ along the pore axis.
The different curves represent results for
particles with fixed radius $R=3\,$nm
and various lengths, namely
$L=6$ nm (black), $L=10$ nm (red), 
$L=18$ nm (blue), 
and $L=26$ nm (green).
(Overall, the forces increase
(in modulus) with particle length.)
The pore radius was $Q=5\,$ nm,
the membrane thickness $H=20\,$ nm,
the surface charge density  $\sigma  = -50$ mC/m$^2$,
and the concentration $c_0=10$ mM in panels (a),(c),(e), 
and $c_0=100$ mM in panels (b),(d),(f). 
The dielectric constants in (\ref{8}) were 
$\epsilon_w = 80$ and $\epsilon_m=\epsilon_p = 5$.
}
\end{figure}

Fig. \ref{fig4} exemplifies the numerically obtained 
forces for four different particle lengths $L$ 
and two different bulk concentrations $c_0$.
Qualitatively, these results are very similar
for both bulk concentrations, but the absolute values 
of the forces are approximately a decade larger 
for $c_0 = 10$ mM than for $c_0 = 100$ mM.
Further main observations are:

(i) Although the particle is neutral, the dielectric
force $F_e(z)$ does not vanish [Figs. \ref{fig4} (c),(d)].

(ii) The $z$-dependence of this dielectric force 
is quite complicated, in particular if $L\approx H$,
see Fig. \ref{fig2} and the green and blue lines in
Figs. \ref{fig4} (c),(d).

(iii) The hydrodynamic force $F_h(z)$ generated by the
counterion pressure (see below (\ref{17a}))
is much larger than the dielectric force $F_e(z)$.

(iv) The various above predicted ``force plateaux''
are indeed observed. In particular, the total force 
$F(z)$ develops for sufficiently large particle lengths 
$L$ two symmetric, asymptotically $L$-independent 
``plateaux'' $\pm F^m$ [Figs. \ref{fig4} (e),(f)],
for which we derived the analytical approximations 
$F_l^m$ in (\ref{a1}) and $F_g^m$ in Sec. \ref{s4b}.

\begin{figure}
\begin{center}
	\epsfxsize=0.8\columnwidth
	\epsfbox{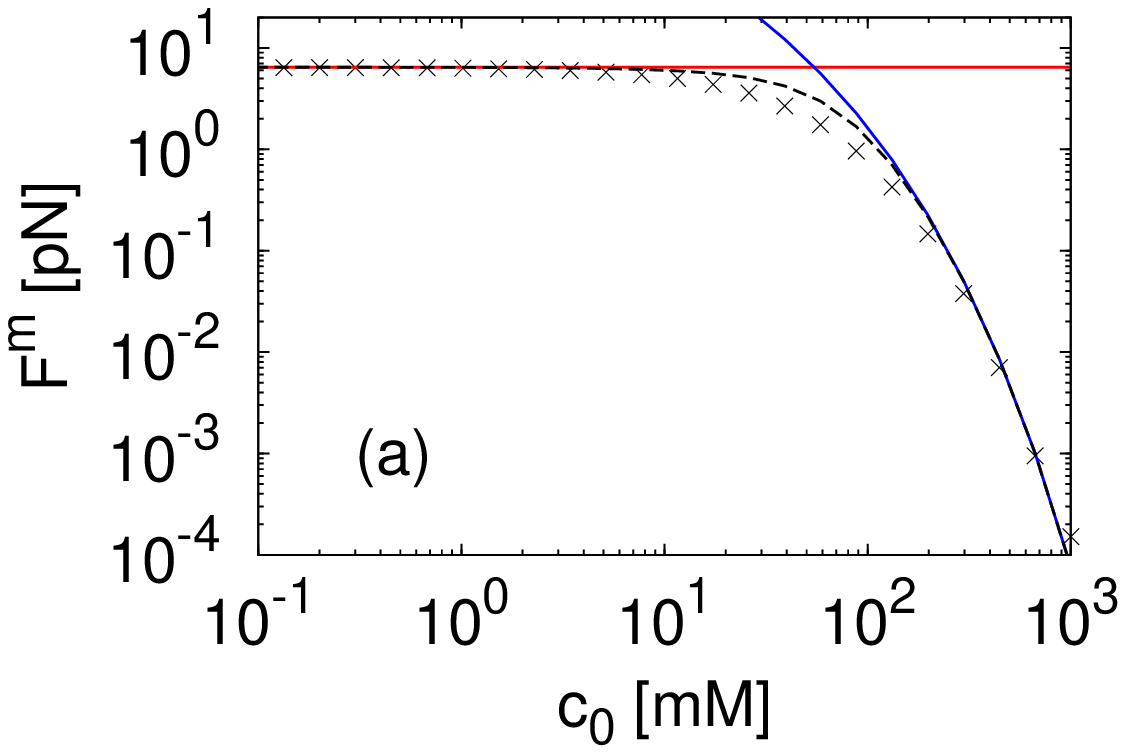}\\
	\epsfxsize=0.8\columnwidth
	\epsfbox{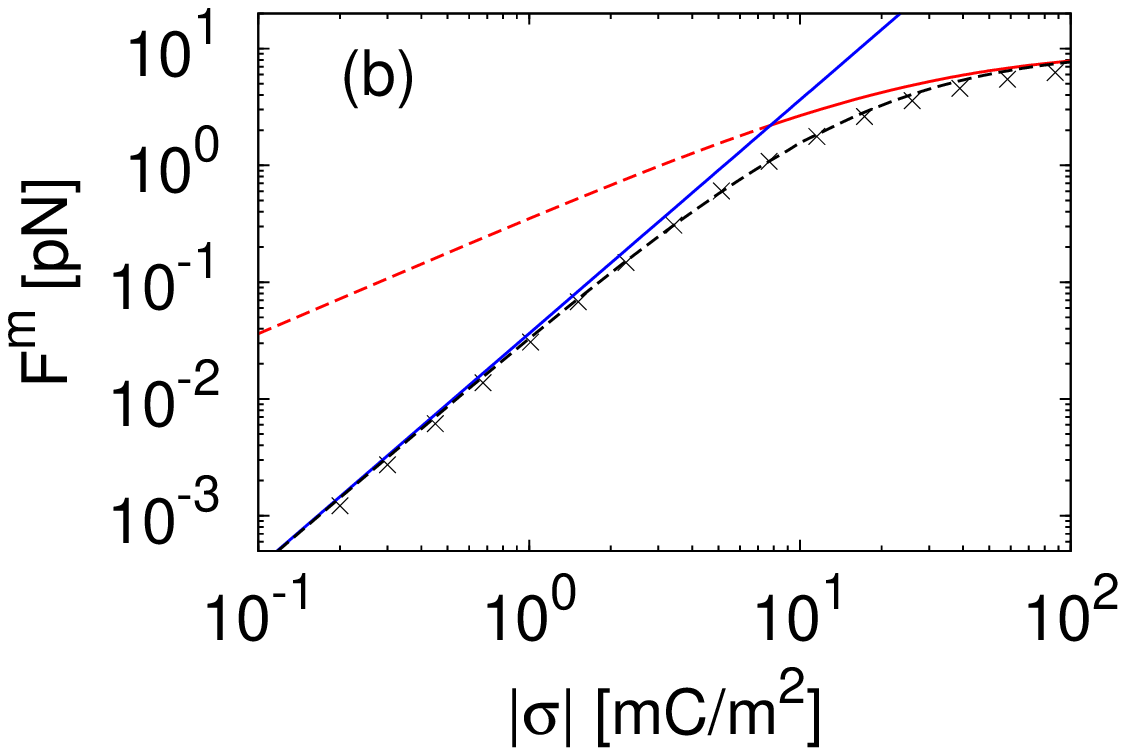}
\end{center}
\caption{\label{fig5}
(Color online)
The maximum (or plateau) force $F^m$ (in units of pN)
for the same system as
in Fig. \ref{fig4} (a),(c),(e),
but now for a particle length of $L=60\,$nm and 
a variable bulk concentration $c_0$ (in units of mM)
in (a) and a variable (negative) surface charge 
density $\sigma=-|\sigma|$ 
(in units of mC/m$^2$) in (b).
Symbols: Numerical solutions.
Red lines (horizontal in (a), 
partially dashed in (b)):
Approximation $F^m_g$ 
for low concentration or high surface charge
from Sec. \ref{s4b}.
Blue lines: Approximation $F^m_h$ from (\ref{a1})
for high concentration or low surface charge.
Dashed: Empirical interpolation (\ref{34m}) 
between these to approximations.
In (b) the red solid line corresponds to 
case ``$+$'' in Eq. (\ref{34a}) and its
dashed continuation to case ``$-$''.
}
\end{figure}

Fig. \ref{fig5} (a) shows the maximum force $F^m$
as a function of the bulk concentration $c_0$ 
for a fixed surface charge density  $\sigma=-50$ mC/m$^2$.
More precisely, the numerical results were obtained
by solving the Poisson-Boltzmann equation (\ref{20a1}) 
for an $L=60$ nm long particle at $z=30$ nm, i.e. with its
lower end at the center of the nanopore.
The numerical solution compares very well with 
the two complementary asymptotic approximations $F_l^m$ 
from (\ref{a1}) and $F_g^m$ from Sec. \ref{s4b}.
In particular, the maximum force $F^m$ becomes (almost) constant 
for low $c_0$, as predicted at the end 
of Sec. \ref{s4b}.

Analogously, in Fig. \ref{fig5} (b) the surface 
charge density was varied while keeping the
concentration $c_0$ fixed at $10\,$mM. 
In particular, $F^m$ indeed scales with
$\sigma^2$ for sufficiently low surface 
charge densities $\sigma$, as predicted by 
the analytical approximation (\ref{a1}) (blue line).

Summarizing Fig. \ref{fig5}, we can say that 
the maximum force is well approximated by $F_l^m$
for low charge densities $\sigma$ and/or 
high concentrations $c_0$, 
and by $F_g^m$ for high charge densities and/or 
low concentrations.
Figs. \ref{fig5} (a),(b) also suggest that
for arbitrary $\sigma$ and $c_0$, at least one
of the two approximations
$F_l^m$ or $F_g^m$ always works reasonably well.
We furthermore observe that $F_l^m\ll F_g^m$ 
in case that $F_l^m$ is a good approximation 
and that $F_g^m\ll F_l^m$ 
in the regime of validity of the approximation
$F_g^m$.
This suggests the following empirical
interpolation formula for the maximum force
\begin{equation}
	\label{34m}
	F_{emp}^m = \left(
		\frac{1}{F_l^m} +
		\frac{1}{F_g^m}
	\right)^{-1}	 \ .
\end{equation}	
The dashed lines in Figs. \ref{fig5} show that this 
approximate ``crossover'' formula indeed works 
remarkably well for arbitrary concentrations 
and charge densities.

\subsection{Potential barriers against entering the pore}
\label{s5c}

\begin{figure}
\begin{center}
	\epsfxsize=0.8\columnwidth
	\epsfbox{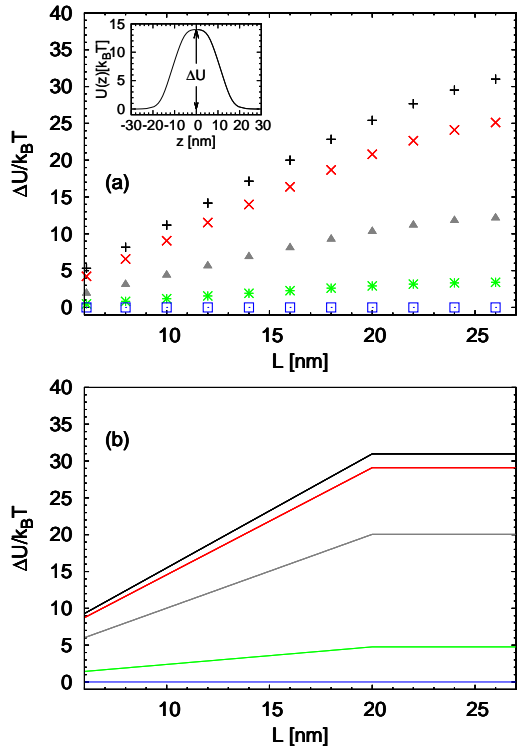}
\end{center}
\caption{\label{fig6}
(Color online)
(a) {\em Inset}: The potential energy $U(z)$ in units of the thermal energy
$k_B T$ versus $z$ (in units of nm), obtained numerically from Eqs. (\ref{15},\ref{16},\ref{17a},\ref{36}),
for a particle length $L=14\,$nm and
all other parameters as in Fig. \ref{fig4} (a), (c), (e).
The potential barrier $\Delta U$ is indicated by the double arrow.
{\em Main figure}: Potential barrier $\Delta U$ 
(in units of $k_B T$)
versus particle length $L$ (in units of nm)
for 5 different bulk concentrations $c_0$, namely
(top-down)
$1$ mM (black), 
$10$ mM (red), 
$40$ mM (gray), 
$100$ mM (green), and 
$1000$ mM (blue).
(b) The corresponding analytical estimates from (\ref{34m}), (\ref{36a}).
}
\end{figure}

Figure \ref{fig6} exemplifies the potential $U(z)$ and, in
particular, the potential barriers $\Delta U$ governing the
pore entrance and translocation by a nanoparticle.
While barriers up to a few $k_B T$ may still be 
surmounted by thermal activation within reasonable 
time-scales, larger barriers practically rule out a 
translocation through the pore
in view of the typical Boltzmann-Arrhenius factors
$\exp(-\Delta U/\kB T)$ by which thermally activated
rate processes are ruled \cite{han90}. 
In conclusion, the translocation of neutral particles 
through nanopores can be greatly suppressed if the 
nanopore walls carry surface charges.

According to Figs. \ref{fig4} and \ref{fig5} 
and their discussion in the main text,
the force $F(z)$ develops two plateaux 
of height $\pm F^m$, whose widths can be
very roughly approximated as $\min\{L,H\}$.
In combination with (\ref{36}), (\ref{34m}) we thus
arrive at the following approximation
for the potential barrier,
\begin{equation}
	\label{36a}
	\Delta U_{emp} = F_{emp}^m \min\{ L,H\} \ .
\end{equation}	
As Fig. \ref{fig6} demonstrates, this simple approximation
reproduced the numerically obtained barriers quite well.
The deviations are mainly rooted in the fact that we 
cannot approximate very well the ``edges'' of the force 
plateaux in Figs. \ref{fig4} (e),(f).

\subsection{Generalizations}
\label{s5d}
So far, we have restricted ourselves to 
particles with radii $R=3\,$nm.
We have seen that, beside other factors, 
the potential barrier for crossing
the pore strongly depends on the ion concentrations.
Above $c_0 \approx 100$ mM the barrier nearly vanishes.
For these salt concentrations, the distance of the particle
from the wall, $Q-R$, is larger than a few times 
the Debye length $\lambda_D$ from (\ref{deb}) 
so that the surface charge
is almost entirely screened by the counterions.
On the other hand, if $Q-R$ is comparable to or smaller than 
the Debye length, the counterion 
pressure within the nanopore will significantly influence the 
translocation dynamics. 
For particle radii other than $R=3$ nm, all effects 
will thus be qualitatively the same as for 
particles with $R=3$ nm, if the pore radius and/or 
the concentration are adapted accordingly.

Next we briefly discuss how the results 
from Sec. \ref{s5b} depend on the surface charge 
density $\sigma$.
As tacitly anticipated in Fig. \ref{fig5}(b), 
all forces are obviously independent of the 
sign of those charges, i.e. they must be 
even functions of $\sigma$.
Moreover, they must vanish in the absence
of any surface charges. 
Hence, the leading order behavior for small 
$\sigma$ will be proportional to $\sigma^2$.
This asymptotics as well the behavior
beyond the small $\sigma$ regime is 
illustrated by Fig. \ref{fig5} (b).
Moreover, we found numerically e.g. for 
$\sigma  = - 20$ mC/m$^2$ 
almost the same shapes of the force curves as
for $\sigma  = -50$ mC/m$^2$ in Fig. \ref{fig4}, 
just their overall amplitudes were rescaled by 
the same factor of $\approx 0.5$ as the corresponding 
maximum force $F^m$ in Fig. \ref{fig5} (b).
A similar behavior is expected for a large
range of other $\sigma$-values.

While the surface charge density of sulfate coated surfaces is,
e.g., quite independent of the solution conditions \cite{beh01,and11},
the surface charge density of silica (SiO$_2$) membranes
has been reported to increase with increasing concentration 
$c_0$ and has typical values between -10 mC/m$^2$ and -100 mC/m$^2$
for $c_0$ between $1\,$mM and $1000\,$mM \cite{beh01,kir04,and11,hoo09,hey05}.
In such a case, a more realistic modeling should
take into account a reduction of the surface charge density
within the pore, compared to the membrane charge density
far from the pore \cite{beh01}. We have conducted preliminary
numerical investigations along these lines, indicating
that the results change only quite insignificantly.

\section{Conclusions}
\label{s6}
We have explored the forces, experienced by
an electrically neutral but in general polarizable 
nanorod in 
an electrolyte solution, which are generated
by a constant surface charge density on 
a membrane with a cylindrical nanopore.
Unless the Debye screening length (\ref{deb}),
quantifying the characteristic extension of
the electric double layer, is much smaller 
than the minimal distance between particle and
pore walls, those forces are quite notable and 
give rise to significant
potential barriers against the particle's 
entrance into the pore.
The dominating contribution is due to the mutual 
repulsion of the counterions which screen the surface charges, 
resulting in an repulsive pressure force on those parts of the
particle which are entering the counterion cloud.
A second contribution is due to the 
combined net effect of all the induced
dipoles in the particle and the ambient fluid.
Under typical experimental conditions those
dielectric forces are, however, much weaker
than the counterion pressure forces.
This is in striking contrast to the extensively
studied opposite case of a charged particle, 
entering a neutral pore 
\cite{par69,jor89,zha07,bon06,kes11}.

What happens if both the pore and the particle 
are charged?
While a systematic exploration of this issue 
goes beyond the scope of our present paper, 
we briefly may point out the main
features of our numerical findings in the 
special case that the pore and the particle 
both carried the same surface charge 
$\sigma  = -50$ mC/m$^2$ (all other parameters
as in Fig. \ref{fig4} (a), (c),(e)):
The forces $F_h(z)$ exhibited
almost the same shapes as those in Fig. \ref{fig4} (a), 
while their amplitudes increased by about a factor 
of five.
The forces $F_e(z)$ did not resemble those
from Fig. \ref{fig4} (b) at all, rather they now
were almost (but not exactly) proportional to 
$F_h(z)$ with proportionality constants close 
to unity.
As a consequence, also the total forces $F(z)$ were similar 
to those from Fig. \ref{fig4} (e), except that 
the amplitudes were larger by about a factor of 
ten.
In particular, these findings cannot be understood
by simply superimposing the cases of an uncharged
particle and of an uncharged pore.
We also note that while $F_h$ can still be associated
with the counterion pressure effects 
(cf. Fig. \ref{fig1}(b) and Eq. (\ref{17a})), 
$F_e$ now comprises not only the dielectric forces but 
also the only partially screened electrostatic
repulsion between the equally charged particle 
and pore walls 
(cf. Fig. \ref{fig1}(a) and Sec. \ref{s3c}).

Regarding potential applications, 
a particularly interesting direction may
be ultrafiltration \cite{bow96}, especially
the design of sieves for uncharged (and
possibly even non-polarizable) nanoparticles,
whose particle sorting characteristics 
can be adjusted by means of the ion 
concentration, see Fig. \ref{fig6}.

\acknowledgments
This work was supported by Deutsche Forschungsgemeinschaft
under SFB 613 and RE1344/8-1 and by the Paderborn 
Center for Parallel Computing.

\section*{Appendix I}
In this Appendix we provide the derivation of 
the approximations (\ref{a1})-(\ref{a3}).

Exploiting $\sinh (x) \approx x$ for $\left|x\right|\lesssim 1$
yields the Poisson-Boltzmann equation (\ref{21}) 
in the Debye-H\"uckel limit
\begin{equation}
	\label{24}
	\frac{1}{r}\dr \left(r\dr \psi_i(r)\right) 
	= \kappa^2\psi_i(r) \ ,
\end{equation} 
where $\kappa := \lambda_D^{-1}$ is
the inverse Debye length, see (\ref{deb}).

By exploiting the above approximation 
$\sinh (x) \approx x$ for $\left|x\right|\lesssim 1$
once again in (\ref{20a}) and the analogous approximation
$\cosh (x) \approx 1+x^2/2$ for $\left|x\right|\lesssim 1$
in (\ref{20b}), we obtain
\begin{eqnarray}
	\rho_i(r)
	& = & - 2Z e_0\NA c_0 \,
	\frac{Ze_0\psi_i(r)}{\kB T}
	\label{20aa}
\\
	p_i(r) & = & \kB T \NA c_0 \left(
	\frac{Ze_0\psi_i(r)}{\kB T}  
	\right)^2 \ .
\label{20bb}
\end{eqnarray}
As a consequence, the integrand in Eq. (\ref{21a})
vanishes and the free energy per unit length simplifies to 
\cite{sha90}
\begin{equation}
	\label{24a}
	g_i = 
	\pi\sigma Q\psi_i(Q)  \ .
\end{equation}

The solutions of Eq. (\ref{24}) with the boundary conditions
discussed below Eq. (\ref{21}) are well known (see e.g. \cite{ric65,gho07})
and are given by (\ref{a2}) and
\begin{equation}
	\label{24c}
	 \psi_2(r) = \frac{\sigma}{\epsilon_0\epsilon_w\kappa} 
	\frac{I_0(\kappa r)K_1(\kappa R)+K_0(\kappa r)I_1(\kappa R)}
		{K_1(\kappa R)I_1(\kappa Q)-I_1(\kappa R)K_1(\kappa Q)} \ ,
\end{equation}
where $I_k$ ($K_k$) is the modified Bessel function of the first (second) 
kind and order $k$.

Introducing (\ref{a2}) and (\ref{24c}) into (\ref{24a}) and 
(\ref{23}) yields (\ref{a1}).
Introducing (\ref{a2}) into (\ref{20bb}) yields (\ref{a3}).

\section*{Appendix II}
In this Appendix we provide the analytical
solution $\psi_2(r)$ of the Poisson-Boltzmann 
equation (\ref{21}) within the 
approximation (\ref{29}), originally derived 
in \cite{phi70,lev75}, and assuming different
functional forms, depending on the model 
parameters:
\begin{equation}
	\label{34a}
	\psi_2(r) = 
	\begin{cases}
	\psi_+(r) \  &\text{if}\ \ln\frac{Q}{R}<1 \  \text{and}\ 
		|\sigma|>\frac{2U_0\epsilon_0\epsilon_w\ln\frac{Q}{R}}{Q\left(1-\ln\frac{Q}{R}\right)}\\
	\psi_-(r) &\text{else}
	\end{cases} \ .
\end{equation}

The function $\psi_+$ is given by
\begin{equation}
	\label{34b}
	\psi_+(r) := \sign (\sigma)U_0\ln\left(
		\frac{-a_3}{\kappa^2r^2\cos^2\left(
			a_4+\frac{1}{2}\sqrt{-a_3}\ln\frac{r}{R}
		\right)}
	\right) 
\end{equation}
where $\kappa$ is the inverse Debye length 
(\ref{deb}) and where $U_0$ is given by (\ref{32}).

The potential $\psi_+$ has to satisfy the boundary conditions
discussed below Eq. (\ref{21}) which fix the
parameters $a_3$ and $a_4$. 
A straightforward calculation shows that $a_3$ is thus given by 
\begin{equation}
	\label{34c}
	a_3 := -4/\tan^2(a_4)
\end{equation}
and that $a_4$ is implicitely given as the solution of
\begin{equation}
	\label{34d}
	\cot (a_4)\tan\left(
		a_4 + \cot(a_4)\ln\frac{Q}{R}
	\right)
	= 1+\frac{Q|\sigma|}{2U_0\epsilon_0\epsilon_w}
\end{equation}
in the interval $(\amin ,\pi/2)$ where
$\amin>0$ is implicitly defined via
\begin{equation}
	\label{34e}	
	\left(\pi/2-\amin\right)\, \tan (\amin)
	= \ln Q/R \ .
\end{equation}

Analogously, $\psi_-$ is given by
\begin{equation}
	\label{34f}
	\psi_-(r) := \sign (\sigma)U_0\ln\left(
		\frac{4a_5a_6(\kappa r)^{\sqrt{a_6}}}
		{(\kappa r)^2\left[
		1-a_5(\kappa r)^{\sqrt{a_6}}
		\right]^2}
	\right)
\end{equation}
with
\begin{equation}
	\label{34g}
	a_5 := \frac{2-\sqrt{a_6}}
		{(2+\sqrt{a_6})(\kappa R)^{\sqrt{a_6}}}
\end{equation}
and $a_6\in(0,4)$ being implicitly defined via
\begin{equation}
	\label{34h}	
	\frac{(a_6-4)\left[
	1-\left(\frac{Q}{R}\right)^{\sqrt{a_6}}
	\right]}
	{(2+\sqrt{a_6})-(2-\sqrt{a_6})\left(\frac{Q}{R}\right)^{\sqrt{a_6}}}
	=
	\frac{|\sigma |Q}{U_0\epsilon_0\epsilon_w} \ .
\end{equation}

\section*{Appendix III}
In this Appendix we show that the approximation $F_g^m$
from Sec. \ref{s4b} is independent of the bulk concentration 
$c_0$.

Introducing (\ref{30}) into 
(\ref{34i},\ref{31}) it follows that 
both $p_1(r)$ and $\rho_1(r)$ 
are independent of $c_0$.
The free energy per unit length $g_1$
thus follows with (\ref{21a},\ref{30}) as
\begin{equation}
	\label{34j}
	g_1 = -\sign(\sigma)\pi U_0 \ln(c_0) \left[
		Q\sigma -\int_0^Qdr\, r\rho(r) 
	\right] + \dots \ ,
\end{equation}
where the dots refer to terms which are 
independent of $c_0$.
Multiplying the integral in Eq. (\ref{34j}) by $2\pi$ yields 
the charge per unit length due to the mobile ions.
Employing the Gauss theorem \cite{jac99} and the boundary conditions discussed
below Eq. (\ref{21}) shows that the integral equals $-Q\sigma$ and hence
\begin{equation}
	\label{34k}
	g_1 = -2\sign(\sigma)\pi U_0 \ln(c_0) 
		Q\sigma  + \dots \ .
\end{equation}
Analogously, it can be shown that also $g_2$ from 
(\ref{21a}) is of the form
\begin{equation}
	\label{34l}
	g_2 = -2\sign(\sigma)\pi U_0 \ln(c_0) 
		Q\sigma  + \dots \ ,
\end{equation}
where the dots in (\ref{34k}) and (\ref{34l}) indicate
in general two different terms, both of which are however
independent of $c_0$. We thus can conclude that the maximum 
force $F_g^m$ from (\ref{23}) is independent of $c_0$ as well.

\end{document}